\documentclass[journal]{IEEEtran}
\usepackage{graphicx}
\usepackage{pythonhighlight}
\usepackage{adjustbox}
\usepackage{hyperref}
\usepackage{xcolor}
\usepackage{soul}
\usepackage{subcaption}

\begin{document}

\title{Distributed Architecture for FPGA-based Superconducting Qubit Control}
\author{\IEEEauthorblockN{
Neelay Fruitwala,$^1$
Gang Huang,$^1$
Yilun Xu,$^1$
Abhi Rajagopala,$^1$
Akel Hashim,$^{1,2}$
Ravi K. Naik, $^{1,2}$
Kasra Nowrouzi,$^{1,2}$
David I. Santiago,$^1$
and Irfan Siddiqi$^{1,2}$
}
\IEEEauthorblockA{\\$^1$Lawrence Berkeley National Laboratory, Berkeley, CA 94720, USA
\\$^2$University of California at Berkeley, Berkeley, CA 94720, USA}}

\maketitle

\begin{abstract}

 Quantum circuits utilizing real time feedback techniques (such as active reset and mid-circuit measurement) are a powerful tool for NISQ-era quantum computing. Such techniques are crucial for implementing error correction protocols, and can reduce the resource requirements of certain quantum algorithms. Realizing these capabilities requires flexible, low-latency classical control. We have developed a custom FPGA-based processor architecture for QubiC, an open source platform for superconducting qubit control. Our architecture is distributed in nature, and consists of a bank of lightweight cores, each configured to control a small (1-3) number of signal generator channels. Each core is capable of executing parameterized control and readout pulses, as well as performing arbitrary control flow based on mid-circuit measurement results. We have also developed a modular compiler stack and domain-specific intermediate representation for programming the processor. Our representation allows users to specify circuits using both gate and pulse-level abstractions, and includes high-level control flow constructs (e.g. if-else blocks and loops). The compiler stack is designed to integrate with quantum software tools and programming languages, such as TrueQ, pyGSTi, and OpenQASM3. In this work, we will detail the design of both the processor and compiler stack, and demonstrate its capabilities with a quantum state teleportation experiment using transmon qubits at the LBNL Advanced Quantum Testbed. 

\end{abstract}

\IEEEpeerreviewmaketitle

\section{Introduction}

\IEEEPARstart{R}{oom} temperature RF control systems have become a critical part of the superconducting quantum computing stack.
With qubit counts in the 10s to 100s, general-purpose RF measurement equipment, such as AWGs (arbitrary waveform generators) combined with discrete RF components, have proven to be overly costly and inefficient for qubit control and measurement. As a result, special-purpose instrumentation has emerged, in both the commercial \cite{de2022qblox, mandelis2022focus, ella2022build} and academic \cite{park2022icarus, stefanazzi2022qick, tholen2022presto, xu2021qubic} realms. These systems integrate pulse sequencing, digital pulse generation, and readout, and are typically built around commercially available FPGAs or SoCs.

The ability to make real-time control decisions based on mid-circuit measurements is becoming an increasingly important part of quantum hardware systems; being a key part of several proposed \cite{deliyannis2022improving} and realized \cite{corcoles2021exploiting} quantum algorithms. For superconducting qubits, with coherence times $\sim 100\ \mu s$, realtime feedback requires a controller with latencies $\sim 100\ ns$. In practice, this means that the feedback control logic must be tightly integrated with the pulse sequencing layer; using an external controller or CPU would significantly increase latency. 

In this work, we present an FPGA-based distributed control architecture which combines pulse sequencing with arbitrary measurement-based control flow. Our design consists of a bank of lightweight, configurable processor cores that are designed to tightly integrate with the puslse generation and signal processing gateware. We also provide a Python/JSON-based intermediate representation for writing and compiling dynamic quantum programs.

 

%\IEEEpubidadjcol

\section{Overall Approach and System Requirements}

The scope of this work includes the pulse sequencing and parameterization layer of the FPGA gateware -- \textit{not} the digital pulse generator modules themselves. We designed this layer to interface with the QubiC 2.0 \cite{xu2023qubic} pulse generation and readout modules; though we believe that our architecture can be adapted to other qubit control systems that use digital pulse synthesis methods. 

We designed our system around the following principles/requirements:
\begin{enumerate}
    \item \textbf{Pulse-centric design}: the primary control primitives are RF control/demodulation pulses; no intrinsic information or assumptions about quantum (unitary) operations being performed. This simplifies the processor core design and instruction set, and makes it straightforward to implement non-standard unitary operations (e.g. optimal control based approaches and certain calibration sequences\cite{werschnik2007quantum}).
    \item \textbf{Low-latency}: Superconducting qubits have coherence times $\sim 100\ \mu s$. This means that for conditional operations based on mid-circuit measurements, we require an end-to-end feedback latency $\sim 100\ ns$. 
    \item \textbf{Lightweight}: Real-time pulse generation places high demand on FPGA logic and memory resources, particularly on high-channel count devices such as the Gen3 Xilinx RFSoC \cite{farley2018all}. So, the pulse-sequencing layer should be as lightweight as possible to accommodate a large number of pulse generators on the same SoC/FPGA. 
    \item \textbf{Flexible}: Superconducting qubit systems have a wide variety of architectures and qubit modalities, each with differing control needs (e.g. readout multiplexing factor, qubit coupler control, and desired instantaneous bandwidth). Our architecture needs to accommodate this variety of pulse generator configurations, and be straightforward to configure at the gateware and software level.
\end{enumerate}

\section{Architecture}

Our architecture consists of a bank of soft processor cores that is responsible for the realtime execution of quantum programs, which involves pulse sequencing, parameterization, and triggering. Each processor core is lightweight, and is designed to interface with a small ($\sim 1 - 5$) number of digital pulse generators that are used for qubit control and readout. This design mirrors the parallelism inherent to quantum circuit execution, which ensures scalability; having a bank of parallel cores (and fixed number of output channels per core) avoids bottlenecks/latency issues that can arise in single-threaded designs as channel count grows. To enable mid-circuit feedforward operations, we also include an extensible ``function processor" module for aggregating and distributing (optionally processed) measurement results to the processor cores. 

\begin{figure}
    \centering
    \includegraphics[scale=0.60, trim={150 40 0 40}, clip]
    {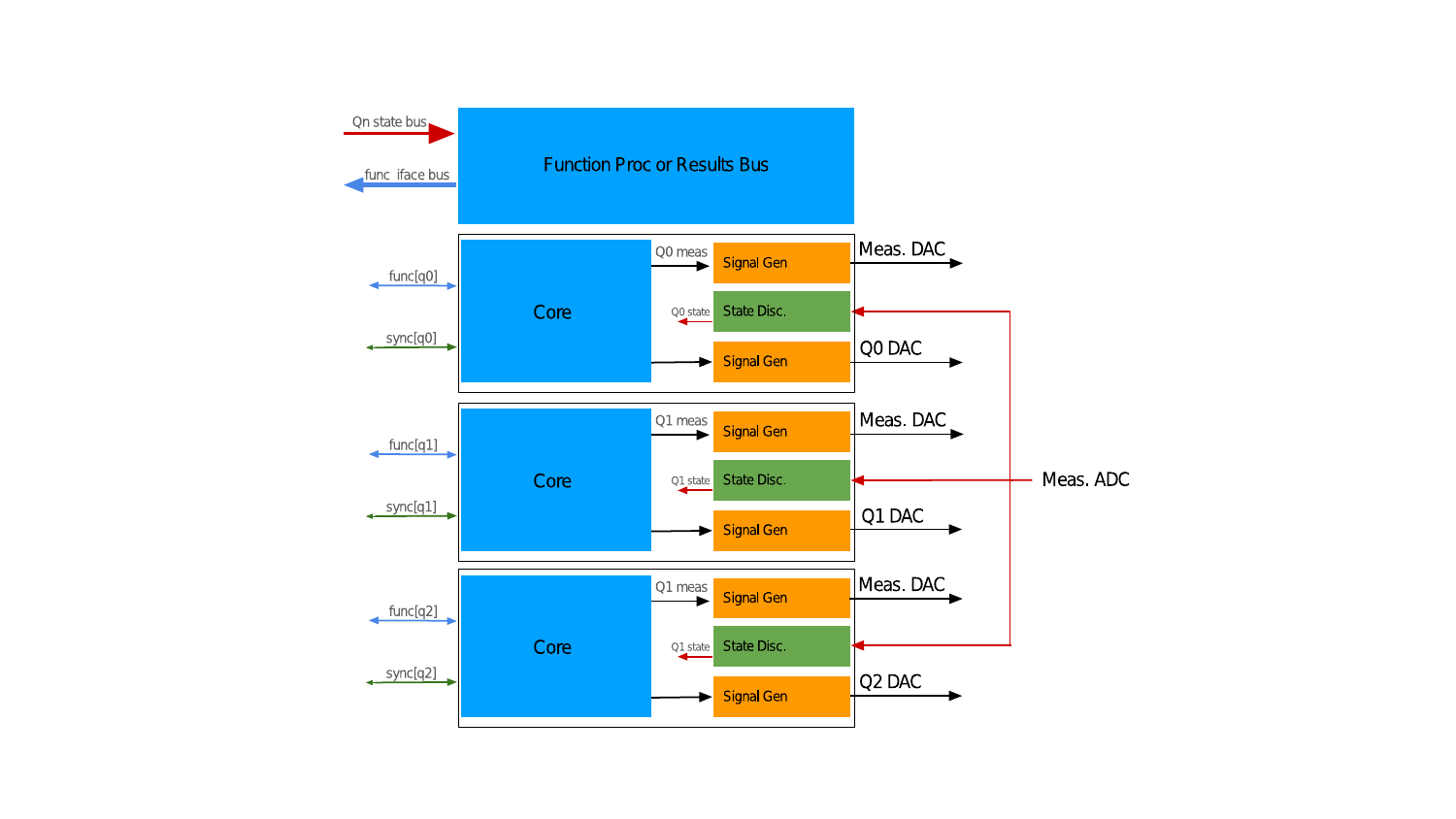}
    \caption{Block diagram of the distributed architecture. In this example, each processor core is responsible for control and readout of a single qubit. Note that the measurement and state-discrimination signal chain exists outside the core, with the results fed directly into the function processor block.}
    \label{fig:full_arch_bd}
\end{figure}

\subsection{Processor Core}

Each processor core implements a custom instruction set architecture (ISA) consisting of pulse commands for real-time control of the associated signal generators, as well as standard arithmetic and control flow instructions for on-the-fly pulse parameterization and the execution of dynamic quantum programs. The full instruction set is detailed in section \ref{sec:instruction_set}. 

\subsubsection{Signal Generator Interface}

Each core is responsible for controlling a small bank of signal generators in real time. This involves both: 1) specifying pulse parameters, such as frequency, phase, and modulation envelope; and 2) triggering the pulse at the correct time. 

QubiC 2.0 \cite{xu2023qubic} uses DDS-based (direct digital synthesis) pulse generation modules, which can synthesize a carrier tone at the provided frequency, phase, and amplitude, and can apply a complex modulation envelope given by a time series of values. In the QubiC 2.0 core $\rightarrow$ signal generator interface, the phase, amplitude, and pulse duration are provided directly via a bus, while the envelope (stored as a series of time-domain values) and frequency (stored as a series of phase offsets per unit time) are pre-allocated in dedicated memory banks, which are configured when uploading the quantum program to the FPGA. It is then the \textit{address} of the envelope/frequency within these buffers that is specified by the processor core.

The processor core $\rightarrow$ signal generator interface consists of the following components: 
\begin{itemize}
    \item Register for storing pulse parameters. Amplitude (16-bit), phase (17-bit), and pulse duration (12-bit) are provided directly, along with pointers to the locations of the modulation envelope (12-bit address) and frequency (9-bit address) in their respective buffers. A configuration word (4-bit) is reserved for miscellaneous parameters.
    \item 1-bit active high pulse trigger (\verb|c_strobe|)
\end{itemize}

The pulse register fields and trigger time are configured by pulse instructions; see section \ref{sec:instruction_set/pulse} for details. 

\subsubsection{Pulse Timing and Synchronization}

All pulse triggers are referenced to an internal counter, which is reset at the beginning of the program. This reset is synchronized across all cores. Additionally, QubiC 2.0 has mechanisms for clock sychronization + synchronized reset across multiple FPGA boards \cite{qubic2}, ensuring that all pulse triggers and reference clocks are synchronized even when cores are distributed across hardware.

\subsubsection{Microarchitecture}
\label{sec:microarch}

\begin{figure*}
    \centering
    \includegraphics[scale=0.65, trim={0 10 0 0}, clip]{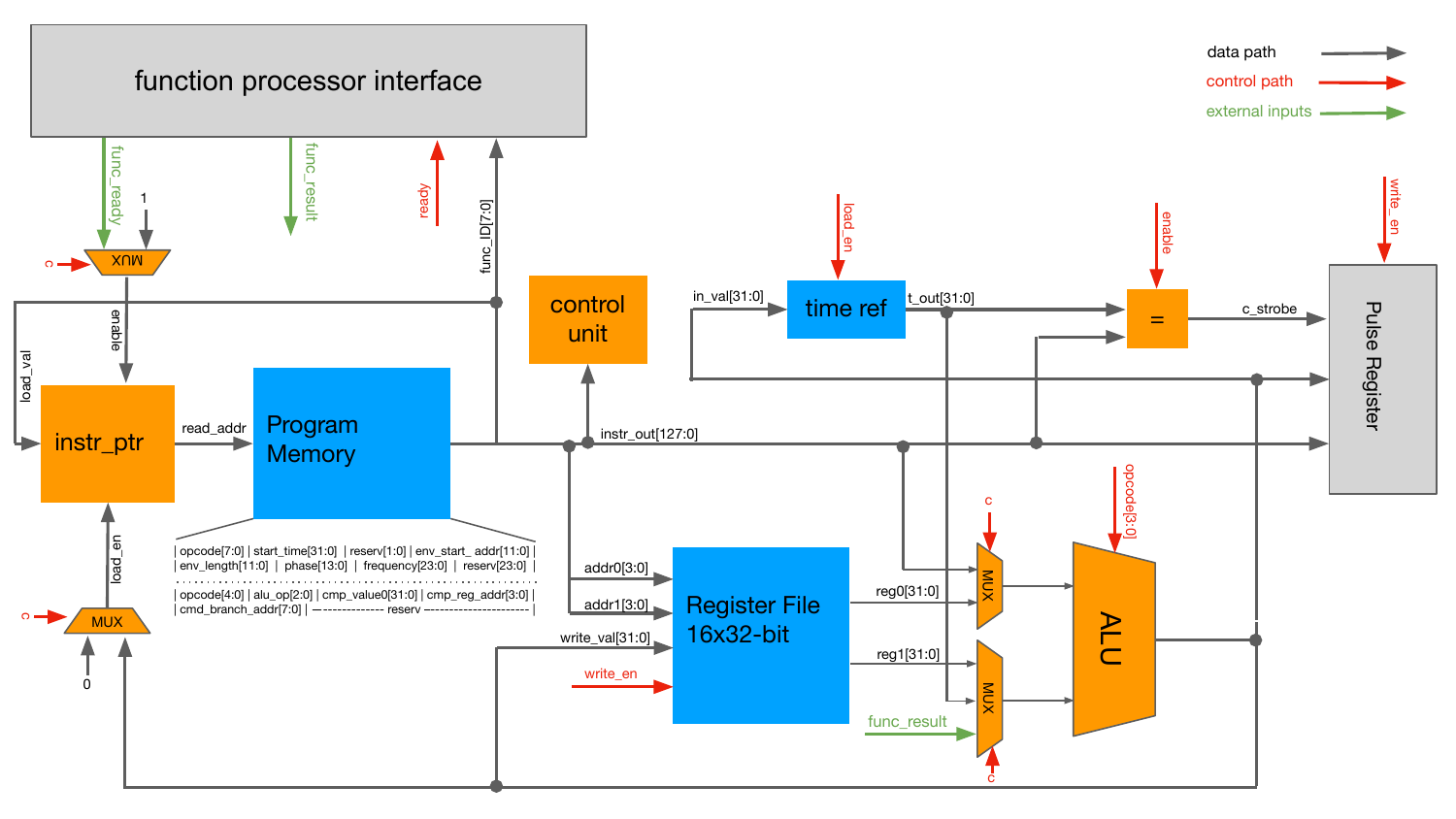}
    \caption{Processor core microarchitecture. Includes a register file, ALU, and instruction pointer for arithmetic and control flow instructions. All pulse triggers are referenced to the \texttt{time\_ref} block, which is a counter that is reset at the beginning of program execution and can be incremented during runtime. All instructions are implemented as 128-bit words. Pulse fields are written to the Pulse Register block, and can be provided by values from the register file and/or instruction immediates. The pulse trigger is given by the \texttt{c\_strobe} signal.}
    \label{fig:microarch_diagram}
\end{figure*}

The processor core microarchitecture is outlined in figure \ref{fig:microarch_diagram}. It is similar to a simple MIPS \cite{kane1992mips} architecture, with a general-purpose 16x 32-bit register bank, 32-bit ALU (arithmetic logic unit), and instruction pointer (or program counter) for interfacing with program memory. For simplicity, the ALU only implements comparison, arithmetic (addition and subtraction), and identity operations. Instructions are implemented using a simple multi-cycle state machine with pipelined instruction fetching. The program memory, pulse interface, and function processor interface are implemented generically in SystemVerilog for portability. We chose an instruction width of 128 bits to accommodate the full 71-bit pulse register, along with the 32-bit pulse start time and other instruction metadata.

\subsection{Function Processor}

Each processor core implements a ``function processor'' interface for connecting to external computational resources. This interface is primarily intended for requesting/receiving (optionally processed) measurement results, although any data/computation with a compatible format can be requested.

The core can request data over this interface by specifying an (implementation specific) 8-bit ID which encodes the type of data to retrieve or computation to perform. Once the data is ready, it is returned as a 32-bit word, along with a ready signal. This request/receive pipeline is triggered by a special instruction, which halts the execution of the core until the resulting data is received. At that point, it can be stored in a register, used as a pulse parameter, or used for a conditional branching decision.

In the current implementation on QubiC 2.0, the function processor interface simply accesses a memory bank containing the most recent state discriminated measurement result from each of the eight qubits driven by the respective FPGA board. This allows any core to request a result from any qubit (provided that it is driven by the same FPGA). Future implementations may extend the function processor to include results from different (synchronized) boards or application-specific measurement decoders.

\section{Instruction Set Architecture}
\label{sec:instruction_set}

Each processor core implements an instruction set consisting of 1) pulse instructions, for parameterizing and triggering pulses; 2) standard register arithmetic and control flow instructions; 3) special-purpose instructions for timing control and interaction with the FPROC interface. In the following section we provide a general overview of the different instruction types; an exhaustive reference can be found in \cite{qubicisa}.

\subsection{Pulse Instructions}
\label{sec:instruction_set/pulse}

In general, there are two different types of pulse instructions: \verb|pulse_write|, which writes to the specified fields of the pulse register, and \verb|pulse_write_trig|, which has all of the functionality of \verb|pulse_write|, but also triggers the pulse at the specified trigger time. The general format for both of these instructions can be found in figure \ref{fig:pulse_instr}.

\subsubsection{\texttt{pulse\_write}}

Pulse register fields can be written to by either an immediate value or a native processor core register (with the exception of a 4-bit configuration word, which must be an immediate). But, only one processor register can be accessed during any given write. So, the \verb|pulse_write| instruction has two additional bits per field: i) \verb|write_enable|, which controls whether to write to that field, and ii) register/immediate select, which controls (if \verb|write_enable| is high) whether the input value comes from the selected register, or the instruction immediate.

\subsubsection{\texttt{pulse\_write\_trig}}

The \verb|pulse_write_trig| instruction adds a 32-bit \verb|start_time| field, which activates the pulse trigger at the provided value, which is in units of FPGA clock cycles since program start and is referenced to an internal counter (figure \ref{fig:microarch_diagram}). Processor core execution is halted until the pulse is triggered.

\subsection{Timing Control}

There are certain situations (for example, when looping over a pulse sequence or waiting for a measurement) where the timing-related behavior of a program must be altered. We provide two instructions for this: the \verb|inc_qclk| instruction, which will increment the time reference by a signed immediate or register value, and the \verb|idle| instruction, which halts execution of the core until the provided timestamp.

\subsection{Arithmetic and Logical Operations}

Register-based arithmetic and boolean operations are performed using \verb|reg_alu| instructions. Supported operations include boolean comparisons ($<$, $>$, $=$), identities, addition, and subtraction, all on 32-bit signed values. Both register-based and instruction immediates are supported. Results are always stored in a register.

\begin{figure*}
\centering
\begin{tabular}{|c|c|c|c|c|c|c|c|}
%\centering
\hline 
127:124 & 123 & 122:120 & 119:88 & 87:84 & 83:68 & 67:52 & 51:0 \\
 \hline 
 opcode & (r/i) &  ALU op & ALU input 0 (r/i) & ALU input 1 & dest reg or jump addr & FPROC ID & 52'b0 \\ \hline

\end{tabular}
\caption{General format for arithmetic and control flow instructions. The instruction type is given by the opcode. Bit 123 (r/i) is used to specify whether ALU input 0 is an instruction immediate or register value from the provided address. The \texttt{inc\_qclk} instruction also follows this format, with only the opcode fields (127:120) and ALU input 0 provided.}
\label{fig:alu_instr}
\end{figure*}

\begin{figure*}
\centering
\resizebox{\textwidth}{!}{
\begin{tabular}{|c|c|c|c|c|c|c|c|c|c|c|c|c|c|}
\hline 
127:120 & 119:116 & 115:114 & 113:90 & 89:88 & 87:71 & 70:69 & 68:60 & 59:58 & 57:42 & 41 & 40:37 & 36:5 & 4:0\\
 \hline 
 opcode & reg addr & env ctrl & env word & phase ctrl & phase word & freq ctrl & freq word & amp ctrl & amp word & cfg en & cfg word & start time & 0 \\ \hline
\end{tabular}}
\caption{General format for \texttt{pulse\_write} and \texttt{pulse\_write\_trig} instructions. Each pulse field (env, phase, amp, and freq) has two control bits; one for write enable and another to select register (from address in 119:116) or instruction immediate. In our implementation, the phase and amplitude are specified directly as scaled values, the frequency is provided as an address, and the envelope word specifies both the start address and envelope length. A 4-bit config word is provided for miscellaneous configuration parameters; this must be provided as an instruction immediate. The \texttt{idle} instruction also follows this format, but the only provided fields are the instruction opcode and the \texttt{start time}, which provides the timestamp after which to resume core execution.}
\label{fig:pulse_instr}
\end{figure*}

\subsection{Control Flow}

Any ALU-based boolean comparison can be used to control a jump instruction, which will set the instruction pointer to an arbitrary location in the program memory. Destination addresses must be instruction immediates. Unconditional jumps are also supported.

\subsection{Function Processor}

Function processor instructions are used to request/receive data over the FPROC interface. These instructions extend ordinary ALU and control flow instructions, but replace one of the fields with the FPROC result. For example, the \verb|jump_fproc| instruction replaces the RHS input of the jump condition with the FPROC result.

% \subsection{Other Instructions}

% Additional instructions include:
% \begin{itemize}
%     \item \texttt{phase_reset}: 
% \end{itemize}

\section{Assembly Language}

We provide a human-readable assembly language that is approximately a one-to-one mapping to the processor core instruction set. The language is formatted as a list of JSON \cite{crockford2017standard} strings, with the assembler and associated infrastructure written in the Python programming language. 

The assembly language instruction fields match those of the instruction set with the following exceptions:
\begin{itemize}
    \item \textbf{Pulse parameters}: all pulse parameters (frequency, amplitude, phase) are provided as floating point values. Frequency and phase are given in SI units, while amplitude is normalized to the DAC full scale. Envelopes are provided as parameterized functions or complex NumPy arrays. Pulse output channels are named, and resolved to gateware/hardware indices during assembly.
    \item \textbf{Register names}: for readablity, register names are provided as strings, and are resolved into indices during assembly. 
    \item \textbf{Register types}: for straightforward pulse parameterization, registers are typed as \verb|amp|, \verb|phase|, or \verb|int|. All operations on \verb|amp| and \verb|phase| type registers are provided in their respective units (\verb|float| in range $[0, 1]$ for \verb|amp| and radians for \verb|phase|), and are converted to the corresponding pulse-field word during assembly. No conversion is performed with \verb|int| type registers.
    \item \textbf{FPROC ID}: function IDs can optionally be specified according to named output channel attributes in the provided channel configuration file. For example, in the program in figure \ref{fig:asm_reset}, the function ID is provided by the \verb|core_id| parameter of the \verb|Q1.rdlo| channel.
\end{itemize}

The assembler takes as input a separate list of instructions for each core, and generates the following outputs: 1) per core program binaries; 2) corresponding set of envelope and frequency buffers. These binaries are stored in a Python dictionary, where they can be loaded by the low-level QubiC driver software into the FPGA BRAM (block-RAM). The assembler is configured using the following:
\begin{itemize}
    \item \textbf{\texttt{ElementConfig} implementation}: \texttt{ElementConfig} is a generic Python class that is implemented separately for each type of firmware signal generator block. It is responsible for converting the provided pulse phases and amplitudes into the correctly formatted words, and computing the frequency and envelope buffers.
    \item \textbf{Channel configuration file}: this file maps named output channels to firmware channel indices. It may also optionally parameterize the implemented \texttt{ElementConfig} class.
\end{itemize}

\begin{figure}
\begin{python}
{('Q1.qdrv', 'Q1.rdrv', 'Q1.rdlo'): [
  {'op': 'phase_reset'},

  # readout drive pulse
  {'op': 'pulse', 'freq': 6.5578e9, 'phase': 0.0,
   'amp': 0.041, 
   'env': {
        'env_func': 'cos_edge_square',
           'paradict': {
                'ramp_fraction': 0.1, 
                'twidth': 1.6e-06}},
   'start_time': 5, 'dest': 'Q1.rdrv'},
   
  # readout demodulation pulse
  {'op': 'pulse', 'freq': 6.5578e9, 'phase': 0.0,
   'amp': 1.0,
   'env': {
        'env_func': 'square',
            'paradict': {
                'phase': 0.0, 'amplitude': 1.0, 
                'twidth': 1.59e-06}},
   'start_time': 325, 'dest': 'Q1.rdlo'},

  # idle to wait for measurement
  {'op': 'idle', 'end_time': 1184},

  # jump instruction; jump to 'true_1' 
  # if measured state is 1
  {'op': 'jump_fproc',
   'in0': 1,
   'alu_op': 'eq',
   'jump_label': 'true_1',
   'func_id': ('Q1.rdlo', 'core_ind')},

  # if state is 0, jump to end
  {'op': 'jump_label', 'dest_label': 'false_1'},
  {'op': 'jump_i', 'jump_label': 'end_1'},

  # if state is 1, play pulse
  {'op': 'jump_label', 'dest_label': 'true_1'},
  {'op': 'pulse',
   'freq': 4.67035e9, 'phase': 0, 'amp': 0.5,
   'env': {
        'env_func': 'DRAG',
        'paradict': {
            'alpha': 0,
            'sigmas': 3,
            'delta': -260.157e3,
        'twidth': 3e-08}},
   'start_time': 1195, 'dest': 'Q1.qdrv'},

  # program end
  {'op': 'jump_label', 'dest_label': 'end_1'},
  {'op': 'done_stb'}],
}
\end{python}
\caption{Example assembly code for single-qubit reset. This program initiates a readout on \texttt{Q1}, then conditionally plays a drive pulse depending on the measurement outcome. The assembly program is formatted as a Python/JSON dictionary, with the program for each processor core keyed by a tuple of channels controlled by that core. In this example, we are only using the qubit \texttt{Q1}, which is controlled by the \texttt{('Q1.qdrv', 'Q1.rdrv', 'Q1.rdlo')} core.}
\label{fig:asm_reset}
\end{figure}

\section{Compiler Tools and Intermediate Representation}

In order to provide users with a high-level format for writing QubiC programs, and to interface with higher-level tools such as TrueQ \cite{trueq}, OpenQASM \cite{cross2022openqasm}, and PyGSTi \cite{nielsen2020probing} we provide a custom intermediate representation (QubiC-IR), along with a set of compiler tools for lowering QubiC-IR to distributed processor assembly. 

We designed QubiC-IR to have the following general attributes:
\begin{enumerate}
    \item \textbf{Multi-level}: In order to provide users with a variety of interfacing options (e.g. native-gate level vs pulse level), QubiC-IR operates at multiple abstraction layers. Only a subset of instructions is directly compilable into distributed processor assembly.
    \item Program flow is \textbf{single-threaded}; the scheduling and compilation tools will parallelize control operations and determine which core(s) need to be targeted by each instruction.
    \item As with the assembly language, QubiC-IR is primarily represented as \textbf{JSON}; IR lowering and compilation is performed using a \textbf{Python API}
\end{enumerate}

The bulk of the compilation is performed in a series of passes that transform the IR. Once the IR has been sufficiently lowered, a final pass will convert it to distributed processor assembly. The compiler flow is customizable; users can both configure individual passes and specify the set of passes to run.

In the following sections, we give an overview of IR instruction types and associated compiler flows. A full reference can be found at \cite{qubicirref}.

\subsection{Control Operations: Gates and Pulses}

QubiC-IR supports a \verb|Pulse| instruction that is largely identical to that of the assembly language. We also support a \verb|Gate| instruction that allows the program to be written at the native quantum gate level, which can then be resolved into pulses by specifying a calibration file containing the pulse parameters associated with each gate.

\subsection{Classical Variables and Arithmetic}

QubiC-IR supports the declaration and manipulation of variables to perform classical computations. Variables are a generalization of assembly language registers; supported operations and allowed datatypes (\verb|int|, \verb|phase|, \verb|amp|) are the same. 

However, unlike registers, a variable can be scoped to multiple processor cores, indicating that the variable declaration itself and any manipulations should be duplicated across the relevant cores as register operations. The scope of any variable is specified by the list of hardware output channels it influences (through either control flow operations or direct pulse parameterization). 

\subsection{Virtual-Z Instructions and Phase Tracking}

In general, virtual-Z gates are implemented by applying a phase offset to any subsequent control pulses at the specified qubit frequency. QubiC-IR supports a \verb|VirtualZ| instruction for this purpose, with two arguments: 1) qubit frequency, and 2) rotation angle (in radians). The qubit frequency can be named (having been previously declared in the program, or defined in the gate calibration file), or anonymous (specified directly using its numerical value).

By default \verb|VirtualZ| instructions are resolved in software; the provided phases are applied directly to the relevant control pulses during compilation. However, hardware (i.e. on-FPGA) resolution is also supported; a \verb|BindPhase| directive can be used to \textit{bind} the phase of all control pulses at a particular frequency to a declared variable. For example, the snippet:
\begin{python}
{'name': 'declare', 'var': 'q0_phase', 
    'dtype': 'phase'},
    
{'name': 'bind_phase', 'var': 'q0_phase', 
    'freq': 'Q0.freq'}
\end{python}
will result in all control pulses with frequency \verb|Q0.freq| to have their \verb|phase| parameter given by the variable \verb|q0_phase|. Hardware phase parameterization is required for certain dynamic circuit operations, such as conditional/repeated application of Z-gates.

\subsection{Control Flow}

QubiC-IR supports both high-level and low-level (assembly-like) control flow. At the high-level, there are two instructions: \verb|BranchVar| and and \verb|Loop|. The \verb|BranchVar| instruction functions as a conditional execution (if/else) statement; the instruction contains an ALU conditional operation to evaluate, and \verb|true| and \verb|false| code blocks which conditionally execute depending on the result of the conditional. The \verb|true| and \verb|false| blocks can contain any valid IR code, including nested control flow. The \verb|Loop| instruction consists of an ALU condition, along with \verb|body| code that executes repeatedly while the condition evaluates to True.

High-level control flow instructions are resolved into lower level \verb|Jump| and \verb|JumpCond| instructions. These are identical to their assembly-level counterparts, except, like ALU arithmetic instructions, they can be scoped (hence duplicated) across multiple processor cores.

After all control flow is lowered to assembly-like control flow, another pass will divide the program into basic blocks, along with the associated control-flow graph (CFG). The CFG is then used by subsequent passes, such as virtual-Z phase resolution and scheduling, to track changes in program state across the full program flow.

\begin{figure}
    \begin{python}
[
    # Since we have conditional z-gates,
    # we need to parameterize the phase of 
    # all Q0 drive pulses with a variable
    {'name': 'declare', 'var': 'q0_phase', 
        'scope': ['Q0'], 'dtype': 'phase'},
    {'name': 'bind_phase', 'var': 'q0_phase', 
        'qubit': 'Q0'},

    # Wait 500 microseconds for qubits to decay
    {'name': 'delay', 't': 500.e-6}
        
    {'name': 'read', 'qubit': ['Q1']},

    # Measurement-based conditional branching 
    # operation. Condition being evaluated 
    # is: `Q1.meas == 1`. Function ID and 
    # associated measurement delays are 
    # resolved by the compiler.
    {'name': 'branch_fproc', 'cond_lhs': 1, 
        'alu_cond': 'eq', 'func_id': 'Q1.meas', 
        
        'true': [
            {'name': 'X90', 'qubit': ['Q0']},
            {'name': 'X90', 'qubit': ['Q0']}
            ],
        'false': [
            {'name': 'virtual_z', 'phase': np.pi, 
                'qubit': 'Q0'}
        ], 
        
        'scope': ['Q0'],},

    # scheduling barrier, then final readout
    {'name': 'barrier', 'qubit':['Q0', 'Q1']},
    {'name': 'read', 'qubit': ['Q0']},
    {'name': 'read', 'qubit': ['Q1']}
]
    \end{python}
\caption{Example program with measurement-based control flow. Q1 is measured, and depending on the outcome of a measurement, either a phase flip or a bit flip is applied to Q0.}
\end{figure}

\subsection{Function Processor}

Special instructions are used to request/receive data over the FPROC interface. As with the assembly language, these instructions extend the normal arithmetic and control flow instructions, replacing the RHS ALU input with the function processor data. 

The QubiC-IR infrastructure can resolve channel names into an assembly-compatible format, and add the appropriate delays to ensure that the processor core has enough time to receive and process the measurement results. This is done using an \verb|FPROCChannelConfig| object and associated compiler pass, which contains a mapping of named FPROC channels to an associated measurement delay and channel ID.

\subsection{Scheduling}

QubiC-IR provides two instructions for specifying timing relationships between gates/pulses -- \verb|Delay| and \verb|Barrier|. The \verb|Delay| instruction delays all subsequent pulses on the specified channels by the provided amount. The \verb|Barrier| instruction is similar to an OpenQASM \cite{cross2022openqasm} barrier; it aligns the start times of the following pulses to be played on the indicated channels. 
%For example, a barrier on channels \verb|Q0.qdrv| and \verb|Q1.qdrv|

The QubiC compiler has a scheduling pass, which assigns trigger timestamps to all \verb|Pulse| instructions. Timestamps are determined by taking into account timing constraints (i.e. delays and barriers), pulse length, and instruction execution time. 

Running the scheduler is optional; users are free to directly provide each pulse with a timestamp. In this case, a linter pass is provided to ensure that the pulse schedule satisfies the execution constraints of the processor core(s) (for example, a pulse cannot be triggered during an ALU or branching operation).

\section{FPGA Implementation}

\begin{figure*}
    \centering
    \includegraphics[scale=0.65, trim={30 150 30 100}, clip]{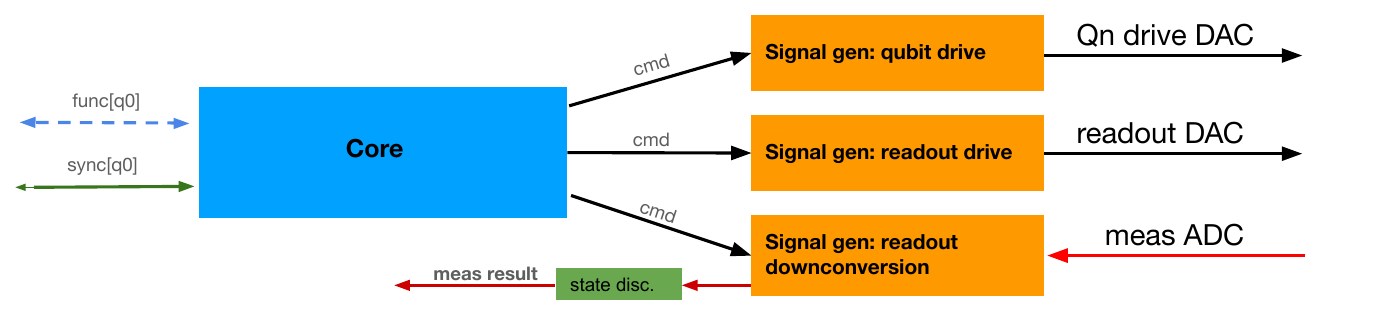}
    \caption{Signal path for a single processor core/qubit for the current QubiC 2.0 implementation. The qubit drive channel goes to a dedicated DAC, while the readout drive (demodulation) channels are connected to a common multiplexed readout DAC (ADC). The state-discriminated measurement result gets sent to the function processor module.}
    \label{fig:single_core_diagram}
\end{figure*}

The processor cores, function processor, and associated interfaces are implemented in Verilog and SystemVerilog. The current implementation is integrated with QubiC 2.0 on the ZCU216 RFSoC platform. Our design is modular; SystemVerilog interfaces are used to connect the processor cores to the QubiC signal generator, memory banks, and measurement modules. 

A variety of QubiC 2.0 implementations exist, ranging from 4x-16x DAC drive channels, and a variety of signal generator and readout multiplexing configurations \cite{xu2023qubic}. In the following sections, we describe a distributed architecture + QubiC 2.0 implementation designed for the AQT Trailblazer QPU (quantum processing unit), which is an 8-qubit superconducting transmon system, with fixed qubit frequency, fixed coupling, and 8x multiplexed readout \cite{kreikebaum2020improving}. 

This implementation has eight distributed processor cores; i.e. one for each qubit. Each core controls three signal generator channels (figure \ref{fig:single_core_diagram}), one for qubit drive (which goes to a dedicated DAC), another for readout drive (which goes to a common multiplexed readout DAC), and another for readout demodulation (which mixes with the readout resonator response tone from the multiplexed readout ADC). All processor cores, signal generator blocks, and readout demodulation are on the same 500 MHz clock domain. All drive and readout DACs are configured to operate at 8 GSPS (gigasamples per second), and the ADCs at 2 GSPS.

\subsection{Resource Utilization}

\begin{figure}
\resizebox{\columnwidth}{!}{
    \begin{tabular}{c|c c c c}
         & LUT & FF & DSP & BRAM \\
         \hline
        processor core &  387 (0.091 \%) & 401 (0.047 \%) & 0 & 2 (0.19 \%)\\
        FPROC & 24 (0.006 \%) & 56 (0.007 \%)  & 0 & 0 
    \end{tabular}}
    \caption{Resource utilization table for a single processor core and full function processor module. Utilization is given as both the absolute number of blocks used and fraction of total utilization for each resource type. The BRAM (block RAM) used by the processor core corresponds to the program memory. Reported values are for the Xilinx ZU49DR FPGA, and were generated using Xilinx Vivado.}
    \label{fig:utilization_table}
\end{figure}

\begin{figure}
    \centering
    \includegraphics[scale=0.57, trim={34 0 15 0}, clip]{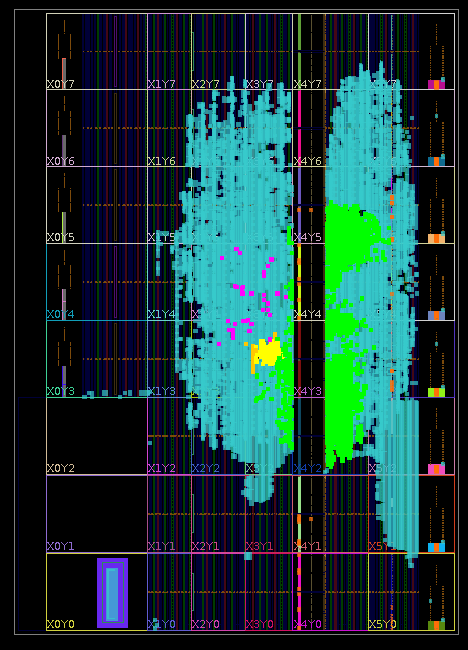}
    \caption{FPGA floorplan of the 8-qubit QubiC 2.0 implementation. Teal-colored cells mark all area utilized by the design (i.e. logic/DSP/memory cells). The highlighted yellow cells mark logic regions used by a single distributed processor core, while green cells mark regions used by the corresponding single-qubit drive and demodulation signal chain. The orange cells highlight block RAM (BRAM) used for processor core program memory. The pink cells mark regions used by the function processor. This floorplan is for the Xilinx ZU49DR FPGA, and was generated using Xilinx Vivado.}
    \label{fig:floorplan}
\end{figure}

The FPGA resource requirements for our implementation are reported in figure \ref{fig:utilization_table} and can be visualized on the floorplan (figure \ref{fig:floorplan}). The logic resource requirements (CLB and DSP) of each processor core are minimal in comparison to the resources required by the corresponding signal processing blocks (i.e. control/readout signal generation and readout demodulation for a single qubit), ensuring that our architecture is unlikely to present a significant scaling bottleneck. The BRAM (block RAM) utilization is largely arbitrary, and depends on the desired circuit depth/program size. In our implementation, single-core BRAM utilization is minimal at 0.2 \%, corresponding to a program memory capable of storing 2048 128-bit instruction words.

\section{Experimental Demonstration: Quantum Teleportation}

\begin{figure}
    \centering
    \includegraphics[scale=0.5, trim={60 50 20 50}, clip]{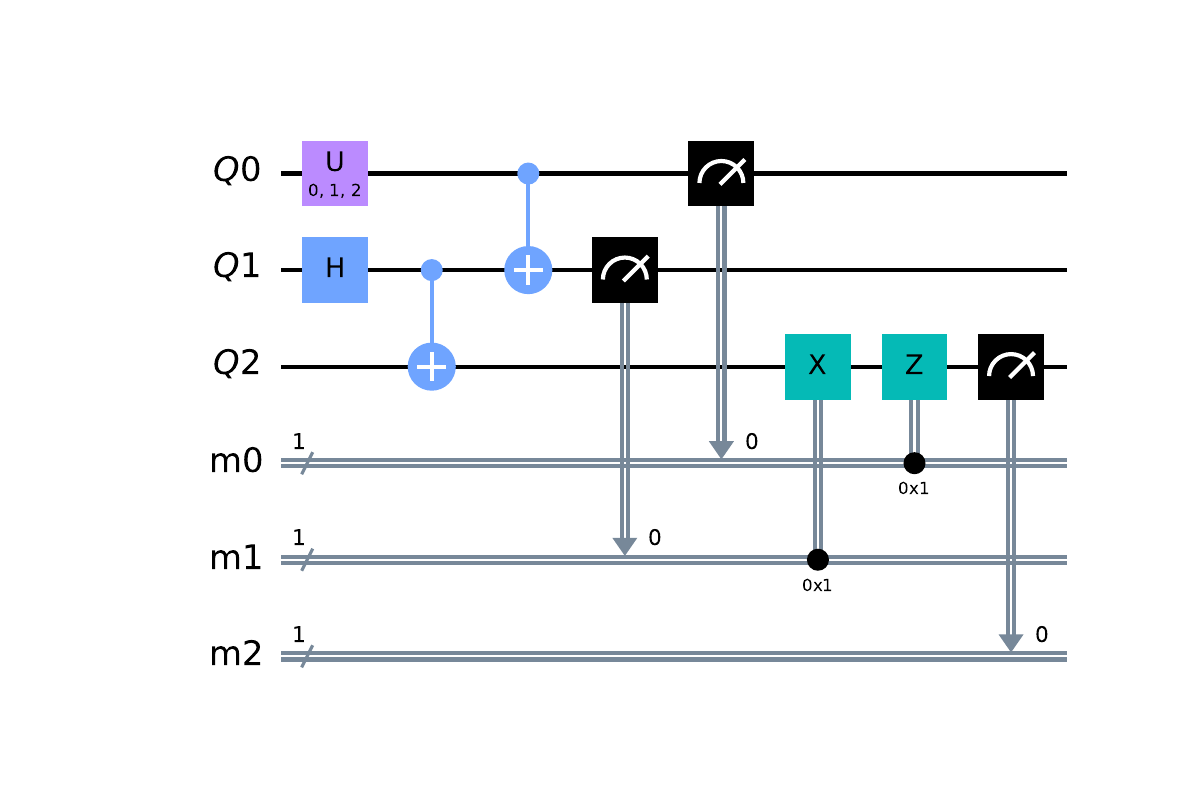}
    \caption{Quantum teleportation circuit. This circuit teleports an arbitrary state encoded on qubit Q0 (prepared by the arbitrary single-qubit rotation U) to qubit Q2. Two mid-circuit measurement-based conditional gates are utilized (the final X and Z gates on Q2).}
    \label{fig:teleport_circuit}
\end{figure}

In order to demonstrate the mid-circuit measurement and feedforward capabilities of our architecture, we performed a quantum state teleportation experiment \cite{bennett1993teleporting}. For this experiment, we used the AQT (Advanced Quantum Testbed) Trailblazer QPU, which has 8 fixed-frequency transmon qubits with linear connectivity \cite{kreikebaum2020improving}. Our teleportation circuit is given in figure \ref{fig:teleport_circuit}; we used the BQSKit compiler \cite{younis2021bqskit} to translate this circuit into the AQT Trailblazer's native gate set (comprised of $X_{90}$, $CZ$, and $virtual$-$Z(\theta)$ gates).

\begin{figure}
    \centering
    \begin{subfigure}{0.25\textwidth}
        \centering
        \includegraphics[scale=0.8, trim={10 20 10 25 }, clip]{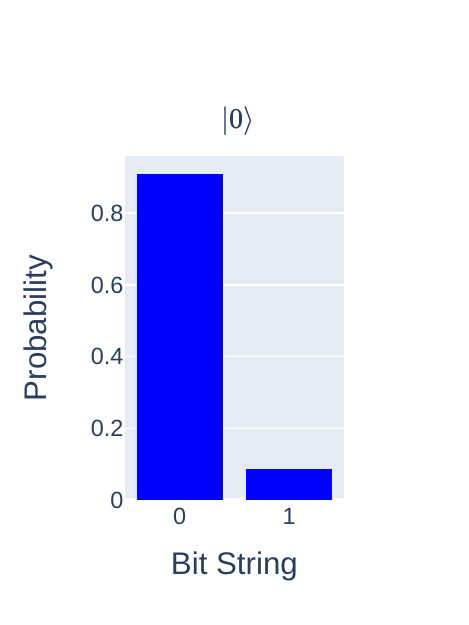}
    \end{subfigure}
    %\hfill
    \begin{subfigure}{0.2\textwidth}
        \centering
        \includegraphics[scale=0.8, trim={35 20 10 25}, clip]{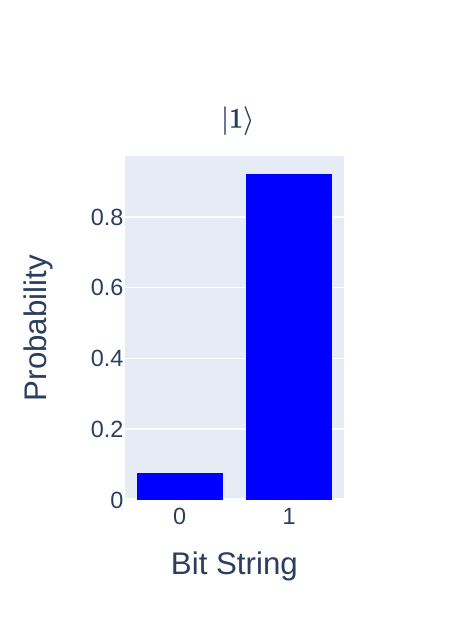}
    \end{subfigure}

    \caption{State teleportation measurement results showing Z-basis measurements of the destination qubit (Q2 from figure \ref{fig:teleport_circuit}) for initial states $|0 \rangle$ and $|1 \rangle$ prepared on Q0. 10,000 shots were collected for each measurement.}

    \label{fig:teleport_01_plots}
\end{figure}

\begin{figure*}
    \centering
    \begin{subfigure}{0.25\textwidth}
        \centering
        \includegraphics[scale=0.8, trim={10 20 10 25 }, clip]{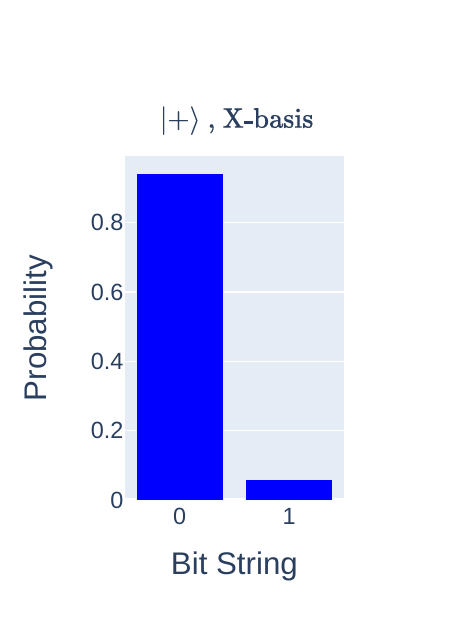}
    \end{subfigure}
    %\hfill
    \begin{subfigure}{0.2\textwidth}
        \centering
        \includegraphics[scale=0.8, trim={35 20 10 25}, clip]{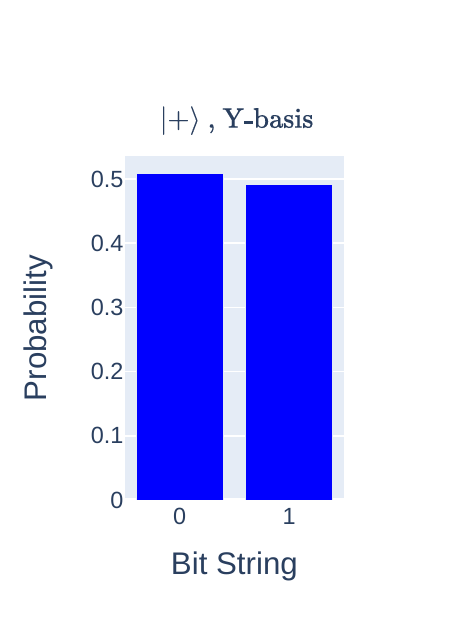}
    \end{subfigure}
    \begin{subfigure}{0.2\textwidth}
        \centering
        \includegraphics[scale=0.8, trim={35 20 10 25}, clip]{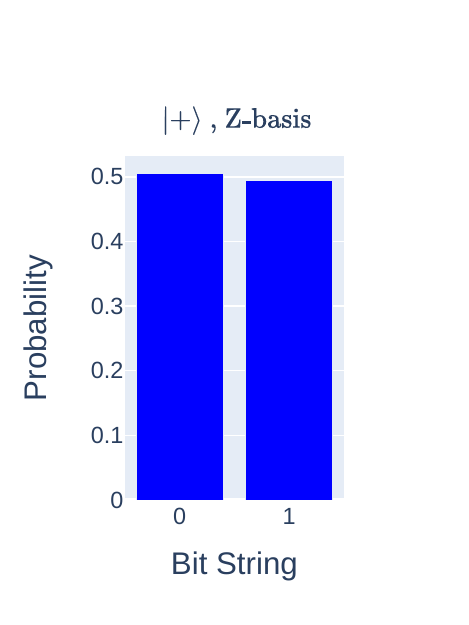}
    \end{subfigure}
    \hfill
    \\
    \begin{subfigure}{0.25\textwidth}
        \centering
        \includegraphics[scale=0.8, trim={10 20 10 25 }, clip]{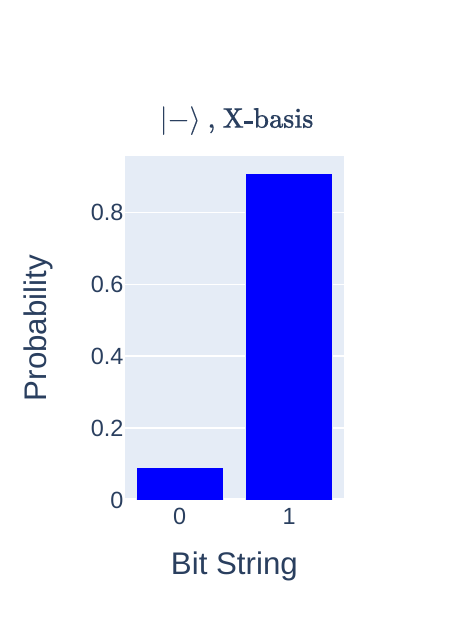}
    \end{subfigure}
    %\hfill
    \begin{subfigure}{0.2\textwidth}
        \centering
        \includegraphics[scale=0.8, trim={35 20 10 25}, clip]{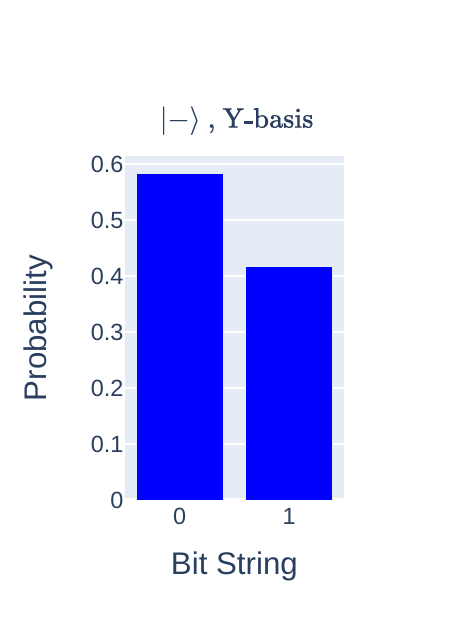}
    \end{subfigure}
    \begin{subfigure}{0.2\textwidth}
        \centering
        \includegraphics[scale=0.8, trim={35 20 10 25}, clip]{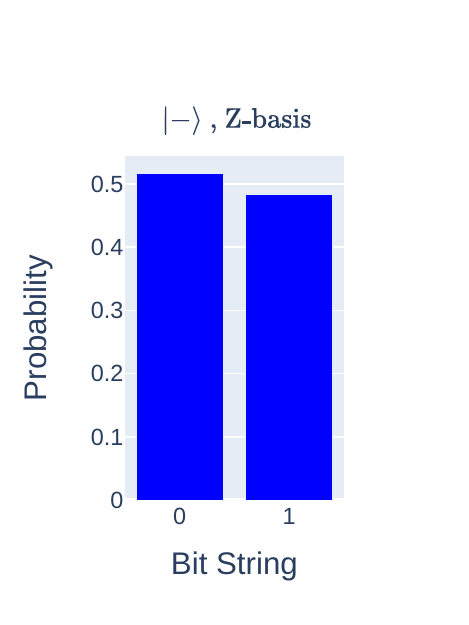}
    \end{subfigure}
    \caption{State teleportation measurement results showing $X$, $Y$, and $Z$ basis measurements of the destination qubit (Q2 from figure \ref{fig:teleport_circuit}) for initial states $|+ \rangle$ and $|- \rangle$ prepared on Q0. 10,000 shots were collected for each measurement.}
    \label{fig:teleport_pm_plots}
\end{figure*}

We performed the teleportation experiment for four different initial states on qubit Q0: $| 0 \rangle$, $| + \rangle$, $| - \rangle$ and $| 1 \rangle$. The corresponding Z-basis measurement results for qubit Q2 are shown in figure \ref{fig:teleport_01_plots} for the $|0 \rangle$ and $|1 \rangle$ initial Q0 states. For the $| + \rangle$ and $| - \rangle$ states, we also performed measurements in the X and Y bases to determine the position of the state vector in the X-Y plane of the Bloch sphere (figure \ref{fig:teleport_pm_plots}).

\begin{figure}
\centering
    \begin{tabular}{c | c | c }
        initial state & basis & expectation
        value \\
        \hline
        \hline
        $|0 \rangle$ & $Z$ & 0.822 (0.008) \\
        \hline
        $|1 \rangle$ & $Z$ & -0.845 (0.008) \\
        \hline
        $ $ & $X$ & 0.884 (0.007) \\
        $|+ \rangle$ & $Y$ & 0.017 (0.015) \\
        $ $ & $Z$ & 0.010 (0.015) \\
        \hline
        $ $ & $X$ & -0.818 (0.008)\\
        $|- \rangle$ & $Y$ & 0.166 (0.015) \\
        $ $ & $Z$ & 0.034 (0.015) \\
    \end{tabular}
    \caption{State teleportation experimental results. Table of expectation values for the destination qubit Q2 for different Q0 initial preparation states and measurement bases.}
    \label{fig:teleport_ev_table}
\end{figure}

A table of measured expectation values for the destination qubit Q2 is given in figure \ref{fig:teleport_ev_table}. For the $| 0 \rangle$, $| + \rangle$, and $| 1 \rangle$ states, the measured expectation values show good agreement with theoretical predictions, indicating successful teleportation of the quantum state from Q0 to Q2. For the $| - \rangle$ state, the measured expectation values show partial agreement, with significant deviation in the Y-basis. This discrepancy is likely due to a combination of dephasing and readout-induced crosstalk, which are known issues when implementing mid-circuit feedforward operations \cite{hashim2024efficient, gambetta2006qubit, mitchell2021hardware}.

% QubiC-IR supports a \verb|VirtualZ| instruction for performing single-qubit Z-gates with arbitrary rotation angles. This is done by applying phase offsets to subsequent control pulses on the specified qubit frequency.

% Assembly language programs are written as a separate list of instructions for each core, and provided to the assembler 

\section{Conclusion}

We have developed an open source FPGA-based architecture for superconducting qubit control and measurement. Our architecture supports the execution of dynamic circuits, including mid-circuit measurement and feedforward, and realtime parameter updates. We also provide a modular compiler stack and intermediate representation that supports a variety of abstraction levels and can integrate with standard quantum programming tools.

Our architecture is deployed on the QubiC 2.0 \cite{xu2023qubic} system, which currently uses the Xilinx ZCU216 RFSoC evaluation board, and has been used to control the 8-qubit Trailblazer QPU at the LBNL AQT. In addition to the state teleportation demonstration presented in this paper, our system has enabled the demonstration of novel scientific results, including randomized compiling for mid-circuit measurement \cite{hashim2024quasiprobabilistic}, and measurement-based entanglement generation \cite{hashim2024efficient}.

Our design and compiler stack is fully open source, and can be found on Gitlab: \url{https://gitlab.com/LBL-QubiC/distributed_processor}.

\appendices

\section*{Acknowledgment}

This work was supported by the U.S. Department of Energy, Office of Science, Advanced Scientific Computing Research Testbeds for Science program, the National Quantum Information Science Research Centers Quantum Systems Accelerator, and the High Energy Physics QUANTISED program under Contract No. DE-AC02-05CH11231.

\ifCLASSOPTIONcaptionsoff
  \newpage
\fi



%\bibliographystyle{IEEEtran}
\bibliographystyle{IEEEtran}
\bibliography{references}


% \end{IEEEbiography}


% % insert where needed to balance the two columns on the last page with
% %\newpage


% % You can push biographies down or up by placing
% % a \vfill before or after them. The appropriate
% % use of \vfill depends on what kind of text is
% % on the last page and whether or not the columns
% % are being equalized.

% %\vfill

% % Can be used to pull up biographies so that the bottom of the last one

\end{document}